\documentstyle[aps,prl,multicol]{revtex}
\tighten
\input epsf

\newcommand{\E}{{\rm e}}
\newcommand{\D}{{\rm d}}
\newcommand{\I}{{\rm i}}
\newcommand{\Tr}{{\rm Tr}}
\renewcommand{\Re}{{\rm Re}}
\renewcommand{\Im}{{\rm Im}}
\newcommand{\sign}{{\rm sign}}

\begin{document}
\draft

\title{Andreev Tunneling in Strongly Interacting Quantum Dots}
\author{R. Raimondi$^{a}$ and  P. Schwab$^{b}$ }
\address{ $^{a}$  Istituto Nazionale per la Fisica della Materia e 
Dipartimento di Fisica E. Amaldi, Universit\`a
 di Roma Tre, Via della Vasca Navale 84, I-00146 Roma, Italy}
\address{
$^{b}$ Dipartimento di Fisica, Universit\'a degli Studi di Roma ``La
Sapienza'', Piazzale Aldo Moro 2, I-00185 Roma, Italy}

\date{\today}
\maketitle

\begin{abstract}
We review recent work on resonant Andreev tunneling through a 
strongly interacting
quantum dot connected to a normal and to a superconducting lead.
We derive a general expression for the current flowing in the structure
and discuss  the linear and non-linear transport in the nonperturbative
regime. New effects associated to the Kondo resonance combined with
the two-particle tunneling arise. The Kondo anomaly in the $I-V$
characteristics depends on the relative size of the gap energy and the
Kondo temperature.
\end{abstract}


\begin{multicols}{2}
\section{Introduction}
In recent years electrical transport through  confined regions 
has seen  impressive theoretical and experimental
activity. Several well known phenomena   have received renewed 
attention due to the present possibility
of studying them in novel and more controllable situations.
For instance, Andreev scattering (Andreev 1964), according to which
a particle-like excitation impinging on a normal metal-superconductor interface is
reflected back as a hole-like excitation, has been shown to be the key
mechanism controlling transport in a variety of hybrid mesoscopic
superconducting devices. (For a review see, for example, Hekking, Sch\"on, and
Averin 1994, Beenakker 1995, Lambert and Raimondi 1998).
At the same time, electron-electron interaction  in small confined
regions, or quantum dots (QDs), has been shown to lead to the so-called
Coulomb blockade of electrical transport. This occurs at low 
temperatures, when the Fermi energy of the contacts 
falls in the gap between the ground state energies of the dot  
with $N$ and $N+1$ electrons (Kastner 1992). 
However, a QD attached to 
metallic leads resembles an impurity level in a metal. As a consequence, even 
in the Coulomb blockade regime,
transport will occur due to the Kondo effect (Glazman and Raikh 1988, Ng 1988). 
This is due to the formation of a 
spin singlet between the impurity level and the conduction electrons, which
gives rise to a quasiparticle peak at the Fermi energy in the dot 
spectral function.  This suggestion  has been explored theoretically by
several  groups (Meir, Wingreen, and Lee 1991,1993, Ng 1993, Levy Yeyati,
Martin-Rodero, and Flores 1993, Schoeller and Sch\"on 1994,
Hettler and Schoeller 1995).
This has lead to the prediction of a
zero-bias anomaly in the current voltage characteristics 
and an increase of the linear conductance 
in the Coulomb blockade regime
for decreasing temperature.
Such phenomena have  indeed been observed in different QD systems
(Ralph and Buhrman 1994, Goldhaber-Gordon et al. 1998, Cronenwett et al. 1998).

What will happen if the QD 
is coupled to a normal and a superconducting
lead as shown schematically in Fig.\ref{fig_bag_dot}?
Does the zero-bias anomaly observed in the N-QD-N case
survive in the N-QD-S case? Such a problem has
been investigated recently by various groups
(Fazio and Raimondi 1998, K. Kang 1999, Schwab and Raimondi 1999,
Sun, Wang, and Lin 1999, 
Clerk, Ambegaokar, Hershfield 1999).
 In this paper we review mainly the work done in Refs. 
 (Fazio and Raimondi 1998) and  (Schwab and Raimondi 1999).

The problem one is faced with has all the difficulty of the original
Kondo problem. 
It is well known that in the limits of high and low temperatures (here
high and low are with respect to the Kondo temperature),  qualitatively correct
results are obtained by means of different techniques.
Such an attitude we adopt here, having in mind to analyze 
the interplay between Andreev scattering and
Kondo physics.  We use a simple equation-of-motion approach
in the perturbative regime above the Kondo temperature. This corresponds to
taking into account only the leading logarithmic corrections in the
renormalization group sense. The cross-over behaviour to low 
temperature may be rather complicated, but as  happens in the
standard Kondo problem, we expect that the extreme low-temperature phase 
can be described in simple terms. Such a description is obtained by
means of the slave-boson technique in a mean-field approximation.
In the high temperature regime,    we find 
a suppressed Andreev current at low bias voltage due to the 
competition between the Coulomb energy and the superconducting 
proximity effect in the QD.  
In the low-temperature case, in contrast, our 
analysis predicts that the linear conductance
of the N-QD-S system is enhanced as compared to the normal case and
may reach the maximum universal value of
$G_{\rm NS}=4e^2/h$ which is twice the maximum for the N-QD-N system.  

Before entering the technical details, it is also worthwhile
  discussing the relevant
energy scales. In investigating Andreev scattering at a 
normal metal-superconductor interface 
one is often interested in voltages and temperatures well below the energy 
scale set by the superconducting gap $\Delta$ and one
could be tempted to take $\Delta \to \infty$ from the outset. 
However, the dot charging energy $U$ (see below)
introduces 
another energy scale into the problem, and the physics is different in 
the two limits $U\gg  \Delta $ and $\Delta \gg U$.
In order to appreciate this point, let us consider the Hamiltonian
of an isolated QD, plus a term which models Andreev scattering in the limit 
$\Delta \to \infty$
by
$H_{\rm A} = t_{\rm A} d^\dagger_\uparrow d^\dagger_\downarrow + c.c $.
For this Hamiltonian the off-diagonal element of the inverse dot 
Green's function reads
$-t_{\rm A}\lbrace 1+U(\epsilon+\epsilon_{\rm d}
+U)/[\epsilon^2-(\epsilon_{\rm d}+U)^2-t_{\rm A}^2]\rbrace$
and vanishes in the $U\to \infty$ limit, {\it i.e.}, 
in the limiting sequence $\Delta \to \infty$ first, $U\to \infty$ after,
the induced superconductivity and hence transport in the dot is completely
 suppressed.
In order to have Andreev scattering in the large $U$ limit two electrons have 
to enter the
superconductor without doubly occupying the QD. 
This can only happen on a time scale of the order $1/\Delta$.
We therefore  concentrate on the limit $U \gg \Delta$ in the following analysis.

Our paper is organized as follows. In Section II we derive a formula
for the current through the dot in the presence of Andreev scattering.
Calculating the current explicitly requires the dot Green's function.
In Sections III and IV we present the equation-of-motion and slave-boson
approaches, respectively. A few conclusions are drawn in section V.

\section{The Andreev current formula}

In this section we define the system under consideration and derive an 
expression for the Andreev current which holds in the presence of 
electron-electron interaction.

The model Hamiltonian for the N-QD-S system is 
\begin{equation}
H=H_{\rm N} +H_{\rm S}+H_{\rm D}+H_{\rm T,N}+H_{\rm T,S}
\label{model}
\end{equation}
where $H_{\rm N}$,  $H_{\rm S}$,  are the Hamiltonians of the normal and the
superconducting leads ($ \Delta $ being the superconducting gap),
\begin{equation}
\label{nlead}
H_{\rm N}=\sum_{{\bf k},\sigma} \epsilon_{\bf k} 
c^{\dagger}_{{\rm N},{\bf k}\sigma} c_{{\rm N} ,{\bf k}\sigma}
\end{equation}
\begin{equation}
\label{slead}
H_{ \rm S}=\sum_{{\bf k},\sigma} \epsilon_{\bf k} c^{\dagger}_{{\rm S},{\bf k}\sigma} 
c_{{\rm S},{\bf k}\sigma}+
\sum_{\bf k} (\Delta c^{\dagger}_{{\rm  S},{\bf k}\uparrow} 
c^{\dagger}_{{\rm S},-{\bf k}\downarrow}
+ c.c.) \;\;\;\;.
\end{equation}
If we restrict ourselves to temperatures and bias voltages much smaller than 
the average level spacing in the QD, then transport occurs through 
a single level. In this case the Hamiltonian of the quantum dot $H_{\rm D}$ reads
\begin{equation}
\label{dot}
H_{\rm D}=\epsilon_{\rm d} d^{\dagger}_{\sigma}d_{\sigma}+Un_{\rm d\uparrow}
n_{\rm d\downarrow} \;\; .
\end{equation}
The  level $\epsilon_{\rm d}$ is
assumed to be spin degenerate and the electron-electron  interaction is 
included through the on-site repulsion $U$ ($ \sim 1-5$K for currently 
fabricated QDs). Experimentally the position of the dot level can be 
modulated by an external gate voltage.
Tunneling between the leads and the dot is described by $H_{\rm T,N}$ and
$H_{\rm T,S}$
\begin{equation}
\label{tunn}
H_{{\rm T},\eta}=
\sum_{{\bf k}\sigma} (V_{\eta} c^{\dagger}_{\eta ,{\bf k}\sigma} d_{\sigma}
+ c.c. )
\end{equation}
where $\eta ={\rm N,S}$ and $V_{\eta}$ is the tunneling amplitude.

The starting point for deriving the current formula is 
$I = -e (\D/\D t) \langle N_{\rm N} \rangle = \I e\left[N_{\rm N}, H \right]$ 
where $N_{\rm N}$ is the 
electron number operator in the normal lead. In the case of a hybrid structure, 
like the one we are considering now, it turns out to be more convenient 
to evaluate the current in the normal lead.
The average current can be rewritten as follows
\begin{equation}
\label{current}
I =
2e{ \Im }\sum_{{\bf k},\sigma}V_{\rm N} \langle c_{{\rm N},{\bf k}\sigma}^{\dagger}
d_{\sigma}\rangle \;\;\; .
\end{equation}
Since we deal with a nonequilibrium situation we work in the framework of the 
Keldysh technique, as employed in the literature (Meir and Wingreen 1992). 
By introducing the Nambu notation
\begin{equation}
\Psi_{{\rm N},{\bf k}} =\left( \begin{array}{c}
 c_{{\rm N},{\bf  k} \uparrow}\\
c^{\dagger}_{{\rm N},-{\bf k}\downarrow}
\end{array}
\right) \;\;\;\;\;\;\;\; 
\phi =\left( \begin{array}{c}
 d_{\uparrow}\\
d^{\dagger}_{\downarrow}
\end{array}
\right),
\end{equation}
the average current in eq.(\ref{current}) requires the evaluation of the 
lesser Green's function 
$
G^{<}_{\alpha \beta ,{\bf k} } (t,t')= \I \langle \Psi^{\dagger}_{\beta,{\rm N},{\bf k}}(t')
\phi_{\alpha}(t) \rangle
$ which, by means of the Dyson equation,
can be expressed in terms of the exact Green's function $G$ of the QD
and the 
free Green's function of the normal lead ${ g}_{{\rm N},{\bf k}}$. 
The lesser component of the Dyson equation reads
\begin{eqnarray}
{\hat G}^{<}_{{\bf k}}(\epsilon) &=& 
{\hat G}^R(\epsilon)
\left( \begin{array}{c c}
V_{\rm N} & 0\\
0 & V^*_{\rm N}
\end{array} \right) 
{\hat g}^<_{{\rm N},{\bf k}}(\epsilon) \nonumber\\
&& +
{\hat G}^<(\epsilon)
\left( \begin{array}{c c}
V_{\rm N} & 0\\
0 & V^*_{\rm N}
\end{array} \right)
{\hat g}^A_{{\rm N} ,{\bf k}}(\epsilon) \;\;\; .
\end{eqnarray}
where ${\hat {G}}^{R(A)}$, ${\hat {G}}^<$
are the retarded (advanced) and
the lesser Green's functions of the dot
(for example ${\hat {G}}^{R}(t)=
-\I \theta (t)\langle\lbrace \phi (t),\phi^{\dagger} (0)\rbrace \rangle$).
Using the relation ${\hat {G}}^<={\hat {G}}^R{\hat {\Sigma}}^<{\hat  {G}}^A$
and the expression for the Green's functions in the normal lead (diagonal in 
Nambu space) $g^R_{{\rm N},{\bf k}}(\epsilon) = 1/(\epsilon - \epsilon_k +\I\eta)$, 
$g^<_{{\rm N} ,{\bf k}}(\epsilon) = 2\pi \I f(\epsilon ) \delta (\epsilon - \epsilon_{\rm k})$
($f(x)$ is the Fermi function), it is possible to rewrite the current 
in the following form
\begin{eqnarray}
\label{current3}
&I = \I e\int_{-\infty}^{\infty}
{{\D\epsilon }\over {2\pi}} \Gamma_{\rm N} \Tr
\lbrace {\hat {\tau}}_z  {\hat {G}}^R(\epsilon) [
 {\hat {\Sigma}}^R(\epsilon) {\hat {f}}_{\rm N} (\epsilon) -&\cr
& {\hat {f}}_{\rm N }(\epsilon){\hat {\Sigma}}^A(\epsilon)+
{\hat {\Sigma}}^<(\epsilon)]{\hat {G}}^A(\epsilon ) \rbrace &
\end{eqnarray}
In the formula above, we introduced the elastic rate 
$\Gamma_{\rm N} (\epsilon ) =2 \pi \sum_{\bf k} |V_{\rm N}|^2
\delta (\epsilon -\epsilon_{\bf k} )$. 
The diagonal matrix ${\hat {f}}_{\rm N}$ has elements
$f_{{\rm N},11}=f(\epsilon -\mu_{\rm N} )$ and
$f_{{\rm N},22}=1-f(-\epsilon -\mu_{\rm N} )$.
The normal electrode is kept  at a chemical potential $\mu_{\rm N}=-eV$ 
while that of the superconductor is fixed to zero.

Up to now no approximations were involved. In order to determine the current 
an expression for the self-energy of the QD should be found. A
determination of $I$ requires both the lesser and retarded parts of the 
self-energy.    
We formulate an ansatz for the lesser Green's function which 
is expressed solely in terms of the retarded one. This ansatz 
automatically guarantees current conservation. 
This generalizes an ansatz 
put forward by (Ng 1996) for two normal leads to the calculation of 
${\hat {\Sigma}}^< (\epsilon)$ in the presence of a superconducting lead.

In order to understand how the ansatz is constructed it is useful to start 
from the noninteracting dot. In this case,
${\hat {\Sigma}}^{<,>}(\epsilon)$ can be computed exactly and it
is expressed in terms of the retarded and advanced self-energies as
\begin{equation}
\label{nonintselfenergy}
{\hat {\Sigma}}^<_0  (\epsilon) =  -\sum_{\eta ={\rm N,S}}\left[
 {\hat {\Sigma}}_{0,\eta}^R(\epsilon) {\hat {f}}_{\eta} (\epsilon) -
 {\hat {f}}_{\eta} (\epsilon){\hat {\Sigma}}_{0,\eta}^A(\epsilon)\right] 
\end{equation} 
\begin{eqnarray}
&{\hat {\Sigma}}^>_0 (\epsilon) =  -\sum_{\eta ={\rm N,S}}
\{{\hat {\Sigma}}_{0,\eta}^R(\epsilon) [{\hat 1}-{\hat {f}}_{\eta}
(\epsilon)]-&\cr 
& [{\hat 1}-{\hat {f}}_{\eta} (\epsilon)]{\hat
{\Sigma}}_{0,\eta}^A(\epsilon)\}.&
\end{eqnarray}
In this case (with no interaction) the nonequilibrium self-energy 
has the same form as in equilibrium but with the Fermi functions of the 
two leads kept at different chemical potentials.  The idea is to assume that 
even in  the presence of interaction the dependence on the Fermi distribution 
is the same and that both the lesser and the greater functions
depend on a single function ${\hat {A}}$
such that
\begin{equation}
\label{intself}
{\hat {\Sigma}}^< ={\hat {\Sigma}}^<_0 {\hat {A}}
\;\;\;\;\;\; , \;\;\;\;\;\;
{\hat {\Sigma}}^> ={\hat {\Sigma}}^>_0 {\hat {A}}
.\end{equation}
The matrix ${\hat {A}}$ is determined by the condition
\begin{equation}
{\hat {\Sigma}}^<-{\hat {\Sigma}}^>={\hat {\Sigma}}^R-{\hat {\Sigma}}^A
,\end{equation}
which is a general property of the Keldysh Green's functions.
 As already
mentioned, this ansatz  leads to current conservation. Moreover it 
is exact both in the non-interacting limit, $U=0$, and in 
absence of  superconductivity, $\Delta =0$.
As a result
\begin{equation}
\label{intselfapp}
{\hat {\Sigma}}^<={\hat {\Sigma}}^<_0 ({\hat {\Sigma}}^R_0 -{\hat {\Sigma}}^A_0)
^{-1}
({\hat {\Sigma}}^R-{\hat {\Sigma}}^A).
\end{equation}
Eq.(\ref{intselfapp}) allows us to evaluate the expression
of the current (\ref{current3}), once we know the retarded
Green's function of the dot.
The  expression for the current can be greatly simplified
in the relevant limit  $U,\Delta \gg k_BT,e V$.
In this case  the  non-interacting self-energy due to
the superconducting lead ${\hat {\Sigma}}^{R(A)}_{0,{\rm S}}$ is real and purely
off-diagonal whereas   that due to the normal lead,
${\hat {\Sigma}}^{R(A)}_{0,{\rm N}}$, is diagonal.
Using these properties of the self-energy  and substituting 
 eq.(\ref{intselfapp}) in eq.(\ref{current3}) 
we obtain the  following form for the
Andreev current through a QD.
\begin{equation}
\label{current4}
I =  \I e\int_{-\infty}^{\infty}
{{\D\epsilon }\over {2\pi}} \Gamma_{\rm N} \nonumber \\
 {\Tr} \left\{ {\hat {\tau}}_z  {\hat G}^R(\epsilon)\left[
 {\hat {\Sigma}}^R(\epsilon), {\hat {f}}_{\rm N} (\epsilon) \right]
 {\hat G}^A(\epsilon ) \right\} \;\;\; .
\end{equation}
This equation generalizes  the formula obtained by
 (Beenakker 1992, Claughton, Leadbeater, and Lambert 1995)
for the non-interacting case to the case
of a strongly interacting dot .

In the following sections we apply this formula to the study of the $I-V$ 
curves of a N-QD-S system.

\section{Equation of motion}
Our aim is to find an equation for the dot's Green's function
$\hat{G}(t)=-\I\langle T\left[ \phi (t)\phi (0)\right] \rangle$. 
The retarded and advanced Green's functions necessary for the evaluation
of the current can easily be obtained from the time-ordered one.
In this section we will describe the equation-of-motion approach.
It is useful to introduce quasiparticle operators
by performing a Bogolubov transformation
\begin{eqnarray}
\label{eom2a}
\gamma _{\eta ,{\bf k}\sigma  } &=&
u_{\eta ,{\bf k} }c_{\eta ,{\bf k}\sigma  }+\sign(\sigma )
\E^{-\I \varphi }v_{\eta ,{\bf k}}c_{\eta ,-{\bf k}\overline{\sigma } }^{\dagger }\\
\label{eom2b}
\gamma _{\eta ,-{\bf k}\overline{\sigma } }^{\dagger }
&=&-\sign(\sigma )\E^{\I\varphi }v_{\eta ,{\bf k} }c_{\eta ,{\bf k}\sigma}
+u_{\eta ,{\bf k} }c_{\eta ,-{\bf k}\overline{\sigma }}^{\dagger }
\end{eqnarray}
where $u^{2}_{\eta ,{\bf k} }(v^{2}_{\eta ,{\bf k} })
=\left(1 \pm \epsilon_{\bf k}/E_{\eta, \bf k}\right)/2$ , 
\( E_{\eta,\bf k}=\sqrt{\epsilon_{\bf k}^{2}
+\mid \Delta _{\eta }\mid ^{2}} \), 
and \( \Delta _{\eta }=\E^{\I\varphi }\mid \Delta _{\eta }\mid  \).
In the normal lead we have \( \Delta _{\rm N}=0 \)
and the Bogolubov transformation reduces to the identity transformation.
The hopping term becomes in the quasiparticle basis 
\begin{equation}
\label{eom3}
\hat{T}_{\eta ,{\bf k} }=\left( \begin{array}{cc}
u_{\eta ,{\bf k} }V_{\eta } & 
-\E^{\I\varphi }v_{\eta ,{\bf k} }V^{*}_{\eta }\\
-\E^{\I\varphi }v_{\eta ,{\bf k} }V_{\eta } 
& -u_{\eta ,{\bf k} }V^{*}_{\eta }
\end{array}
\right) .
\end{equation}
The Nambu
formalism is introduced for the $\gamma$  operator via
$\tilde\Psi_{\eta,{\bf k}}=
(\gamma_{\eta,{\bf k}\uparrow }, 
\gamma^{\dagger}_{{\eta} ,-{\bf k}\downarrow})$.
We start by writing down
the equation of motion for the operators $\phi$ and $\tilde\Psi$:
\begin{eqnarray}
\label{eom6}
\I \frac{\partial }{\partial t}\phi &=&\epsilon _{\rm d}\hat{\sigma }_{z}
+U\hat{\sigma }_{z}\Phi 
+\sum _{\eta ,{\bf k} }\hat{T}^{\dagger }_{\eta ,{\bf k}}
\tilde\Psi _{\eta ,{\bf k} }\\
\label{eom7}
\I \frac{\partial }{\partial t}\tilde\Psi _{\eta ,{\bf k}}
&=& E_{\eta, \bf k}\hat{\sigma }_{z}
\tilde\Psi _{\eta ,{\bf k} }+\hat{T}_{\eta ,{\bf k} }\phi. 
\end{eqnarray}
In the equation for $\phi$, the new operator 
$ \Phi =(
d_{\downarrow }^{\dagger }d_{\downarrow }d_{\uparrow }, 
d^{\dagger }_{\downarrow }d^{\dagger }_{\uparrow }d_{\uparrow } )$,
appears because of the interaction. It is straightforward to iterate once 
the equation of motion and get an equation for $\Phi$
\begin{eqnarray}
\label{eom8}
&\I\frac{\partial }{\partial t}\Phi =\left( \epsilon _{\rm d}
+U\right) \hat{\sigma }_{z}\Phi +\sum _{\eta ,{\bf k}}
\left[ \hat{N}\hat{T}_{\eta ,{\bf k} }^{\dagger }\tilde\Psi _{\eta ,{\bf k} }\right.
 &\cr
&\left. +d^{\dagger }_{\downarrow }d_{\uparrow }
\left( \hat{T}_{\eta ,{\bf k} }\right) ^{t}
\left( \tilde\Psi _{\eta ,{\bf k} }^{\dagger }\right) ^{t}\right] &
,\end{eqnarray}
where 
\[
\hat{N}=\left( \begin{array}{cc}
d_{\downarrow }^{\dagger }d_{\downarrow }, 
& d_{\downarrow }d_{\uparrow }\\
d_{\downarrow }^{\dagger }d^{\dagger }_{\uparrow }, 
& d^{\dagger }_{\uparrow }d_{\uparrow }
\end{array}
\right) .\]
Here the superscript \( t \) means the transpose. 
In eq.(\ref{eom8}) again
several new operators have appeared.  In order
to get a closed system of equations one has to truncate the hierarchy
at some point. Following the decoupling procedure used in the
absence of superconducting leads (Meir, Wingreen and Lee 1991), we 
neglect correlations in the leads.
To see how this is achieved, consider the general structure of all
the new operators entering  eq.(\ref{eom8}) for \( \Phi  \). 
They are products of
two \( d \) and one \( \gamma  \) operator. By iterating once more the equation of
motion one obtains new terms which contain two \( \gamma  \) 
and one \( d \) operator.
The approximation consists then in replacing a pair of two such \( \gamma  \)
operators with their statistical average. 
 In such a way all the new operators entering
the eq.(\ref{eom8}) for \( \Phi  \) can be expressed in 
terms of \( \Phi  \)  itself and \( \phi  \). A further
approximation is done by considering the limit of large Coulomb
interaction, {\it i.e.}, \( U\rightarrow \infty  \). 
In this case one can safely neglect, in the
third order of iteration of the equation of motion, all operators
having two creation or two annihilation \( d \) operators, because they
will give rise to terms of the order \( 1/U \). 
We write here  the resulting equations of motion
for \( {\hat G} \) and 
${\hat G}^{II} =
-\I\left\langle T\left[ \Phi (t)\phi (0)\right] \right\rangle$. They are 
\begin{eqnarray}
\label{eom21}
\left( \omega \hat\sigma_0 
-\epsilon _{\rm d}\hat{\sigma }_{z}\right) {\hat G}(\omega )
&=&\hat\sigma_0+\hat{\Sigma }_{0}(\omega )\hat G(\omega )
+U\hat{\sigma }_{z}{\hat G}^{II}(\omega ) \\
\label{eom22}
U\hat{\sigma }_{z}{\hat G}^{II}(\omega )
&=&-\langle \hat{N}\rangle 
+\hat{\Sigma }_{I}(\omega ){\hat G}(\omega )
\end{eqnarray}
and the equation for the single particle Green's
function becomes
\begin{equation}
\label{eom23}
\left( \omega \hat\sigma_0 -\epsilon _{\rm d}\hat{\sigma }_{z}
-\hat{\sigma }(\omega )\right) 
{\hat G}(\omega )
=\hat\sigma_0-\langle \hat{N}\rangle ,
\end{equation}
where $\hat{\sigma }=\hat{\Sigma }_{0}+\hat{\Sigma }_{I}$.
The noninteracting self-energy \( \hat{\Sigma }_{0} \) is  obtained as
\begin{equation}
\label{eom24}
\hat{\Sigma }_{0}\left( \omega \right) 
=-{\I\over 2} \sum _{\eta }\frac{\Gamma_{\eta }}{\sqrt{\omega ^{2}-\Delta ^{2}_{\eta }}}
\left( \begin{array}{cc}
\omega  & \Delta _{\eta }\\
-\Delta _{\eta } & -\omega 
\end{array}
\right) 
\end{equation}
with 
$\Gamma_\eta = 2 \pi \sum_{\bf k} |V_\eta |^2 
 \delta(\epsilon - \epsilon_{\bf k}).$
For the calculation of the interacting self-energy 
\( \hat{\Sigma }_{I}\left( \omega \right)  \)
we assume that in the superconducting reservoir 
there are no quasiparticles
present (we consider energies much smaller than the gap \( \Delta  \))
\( \langle
\gamma^{\dagger }_{{\rm S}, {\bf k}\sigma }\gamma_{{\rm S},{\bf k}\sigma }\rangle
=0  \), 
and that in the
normal reservoir 
the quasiparticle population follows the Fermi
distribution 
\( \langle
\gamma^{\dagger}_{{\rm N},{\bf k}\sigma }\gamma_{{\rm N},{\bf k}\sigma }
\rangle
=f\left(  \epsilon_{\bf k} -\mu_{\rm N}\right)  \). 
We then have
\begin{equation}
\label{eom25}
\hat{\Sigma }_{I,{\rm S}}\left( \omega \right) 
=V^{2}_{\rm S}\sum _{\bf k}\left( \begin{array}{cc}
\frac{|v_{{\bf k},{\rm S} }|^2}{\omega -E_{{\bf k},{\rm S}}} 
& \frac{2u_{{\bf k},{\rm S}}v_{{\bf k},{\rm S}}}{\omega -E_{{\bf k},{\rm S}}}\\
-\frac{2u_{{\bf k},{\rm S}} v_{{\bf k},{\rm S}}}{\omega +E_{{\bf k},{\rm S}}}
 & \frac{|v_{{\bf k},{\rm S}}|^2}{\omega +E_{{\bf k},{\rm S}}}
\end{array}
\right) 
\end{equation}

\begin{equation}
\label{eom26}
\hat{\Sigma }_{I,{\rm N}}\left( \omega \right) 
=-V^{2}_{\rm N}\sum _{\bf k}\left( \begin{array}{cc}
\frac{f( \epsilon_{\bf k} ) }{\omega -\epsilon_{\bf k}} & 0\\
0 & \frac{f( \epsilon_{\bf k} ) }{\omega +\epsilon_{\bf k}}
\end{array}
\right) 
.\end{equation}
In what follows, we consider
$W \gg\Delta \gg T,\omega$  with $W$ the bandwidth.
The self-energy ${\Sigma}_{I,{\rm S},11}$ 
is weakly energy dependent  
since $\omega \ll \Delta < E_{{\bf k}, \rm S}$.  
The imaginary part of the diagonal 
elements of the self-energy ${\hat \Sigma}_{\rm S}$ vanishes, so
quasiparticles present in the dot cannot decay by tunneling into 
the superconductor. 
As a result the contribution to the self-energy due to the superconducting 
lead simply shifts the dot level to the new value
$\tilde{\epsilon }_{\rm d} \approx \epsilon_{\rm d} +(\Gamma _{\rm S}/2\pi )
\ln W/\Delta$.
The divergence of the level energy renormalization with $\Delta$ reveals
that the process occurs via a virtual state in which a quasiparticle 
is created in the superconductor. 
Substituting the expression of the QD's Green's function 
in eq.(\ref{current4}) we get the  result
\begin{equation}
\label{current55}
I(V) =\int^{\infty}_{-\infty}~\D\epsilon 
{{f(\epsilon -eV)-f(\epsilon +eV)}\over {2e}} G_{\rm NS}(\epsilon )
\end{equation}
with
\end{multicols} 
\begin{equation}
\label{conductancebis}
G_{\rm NS}(\epsilon )={{4e^2}\over h} 
{{2(\Gamma_{\rm N} \Gamma_{\rm S}(\epsilon ) )^2 
\left[(\Gamma_{1,{\rm N}} +\Gamma_{2,{\rm N}})/2\Gamma_{\rm N})\right]}\over
{\left[4 (\epsilon-\epsilon_1)(\epsilon+\epsilon_2) 
 -\Gamma_{1,{\rm N}}\Gamma_{2,{\rm N}}
 -\Gamma_{\rm S}^2(\epsilon )\right]^2 +4\left[\Gamma_{1,{\rm N}}(\epsilon+\epsilon_2)+
  \Gamma_{2,{\rm N}}(\epsilon-\epsilon_1)\right]^2} }
\end{equation}
\begin{multicols}{2}
\noindent
where 
$\epsilon_{1(2)} =\epsilon_{\rm d} \pm \Re{\hat {\sigma}}_{11(22)}$,  
$\Gamma_{1(2),{\rm N}}=-2\Im{\hat {\sigma}}_{11(22)}$, and 
$\Gamma_S (\epsilon ) =2\Re {\hat {\sigma}}_{12}=
\Gamma_S 2\epsilon / (\pi \Delta )$. 
Notice that the anomalous non-interacting $\Sigma_{0,12}$ and interacting 
$\Sigma_{{\rm I},12}$ self-energies exactly cancel in the zero energy
limit, resulting in a linear behaviour for energies smaller than the
gap\cite{correction}. 
The spectral function $G_{\rm NS}(\epsilon )$ associated with the resonant 
Andreev 
tunneling is plotted in Fig.~\ref{fig2} for various bias voltages.
Several features are worth noticing.  First, two peaks at 
$\pm \tilde{\epsilon}$ are due to particle and hole bare levels.
Note that in the interacting case the bare level energy includes
the renormalization due to the superconducting electrode self-energy
as discussed above. Second, at low temperatures the spectral function
at the Fermi energy is completely suppressed, in contrast to
what happens in N-QD-N case.  Quite remarkably, 
at finite positive (negative) voltages a Kondo peak develops pinned to the 
Fermi level of the normal metal while a small kink develops at negative 
(positive) voltages. At  finite voltages  hole and particle energies differ by 
$2eV$, and while the electron (hole) is on resonance for positive (negative)
voltage, the Andreev reflected hole (electron) is off resonance with respect 
to the shifted Fermi level. 

The differential conductance, for various temperatures,
is shown in Fig.~\ref{fig3}. Lowering the  temperature  a  zero-bias 
anomaly  starts to develop where the conductance
is strongly suppressed at low voltages.
From eqs.(\ref{current55}-\ref{conductancebis}) we conclude that the
linear conductance is roughly proportional to $T^2/\Delta^2$, so that
it seems to be completly suppressed in the zero temperature limit.
However,
the equation-of-motion approach is quantitatively reliable only
above the Kondo temperature. Hence the analysis carried out here applies
to the regime $\Delta > T > T_K$.  
An interesting question to ask is what happens in the opposite
regime when the Kondo temperature dominates, i.e., when 
$T_K > \Delta > T$. This regime cannot be explored by the equation-of-motion
approach. For this reason in the next section  
we will investigate the low temperature regime
by means of the slave-boson technique.

\section{Slave boson mean field approximation}
In this section we extend the analysis of the previous section 
to the extreme low temperature regime. To this end we use slave-boson mean
field theory.
This approach has been successfully applied to the low temperature
properties of a Kondo impurity in the presence of normal
(Barnes 1976, Coleman 1984, Read and Newns 1983, Read 1985)
as well as for superconducting conduction electrons 
(Borkowski and Hirschfeld 1994). 
Despite its simplicity, this method captures the main physical aspects
of the Fermi liquid regime at low temperatures, {\it i.e.}, the formation
of a many-body resonance  at the Fermi energy.
For this reason it presents a convenient framework in which to
study the interplay between Andreev scattering and Coulomb interactions.

Again we consider 
an infinite on-site repulsion $U$, so 
processes where the dot level is doubly occupied  are excluded.
The dot level is represented as
$d^\dagger_\sigma = f^\dagger_\sigma b$, 
where the fermion $f_\sigma$   and the boson $b$ describe the singly occupied 
and  empty  dot states.
Since the dot is either empty or singly occupied, the constraint
$b^\dagger b + \sum_\sigma f^\dagger_\sigma f_\sigma =1$
has to be fulfilled.

In mean field approximation, the operator $b$ is replaced by a $c$-number
$b_0$, and the constraint is fulfilled only on average.
This is achieved by introducing  a chemical potential
$\lambda_0$ for the pseudo particles.
Notice that one ends up
with a non-interacting-like problem with renormalized parameters,
{\it i.e.}, an energy shift for the dot level
$\epsilon_{\rm d} \rightarrow \epsilon_{\rm d} +\lambda_0=\tilde \epsilon_{\rm d}$ 
and a multiplicatively renormalized tunneling amplitude
 $V_\eta \rightarrow b_0 V_\eta$.

We discuss the mean field equations and their solution
first in equilibrium and then generalize to non-equilibrium. 
We start from the impurity part of the free energy, which in 
the presence of both normal and  superconducting leads is given by
\begin{equation} \label{sb2}
F= - T\sum_{\epsilon_n} \Tr \ln [ 
\I \epsilon_n \hat\sigma_0 - \tilde \epsilon_{\rm d} \hat\sigma_z -b_0^2
  \hat \Gamma(\I \epsilon_n) ]
+\lambda_0 b_0^2 +\epsilon_{\rm d} - \mu 
,\end{equation}
where $\epsilon_n$ is a fermionic Matsubara frequency,
${\hat {\sigma}}^i$ are the Pauli matrices, and 
\begin{eqnarray}\label{sb3}
\hat \Gamma(\I\epsilon_n )&= & \sum_{{\bf k}, \eta }
 |V_\eta|^2 \hat \sigma_z \hat g_{\eta ,{\bf k}}(\I \epsilon_n) \hat \sigma_z
\end{eqnarray}
with $\hat g_{\eta ,{\bf k}}$ being the Green's function of the
lead $\eta$. 

By minimizing the free energy with respect to $\lambda_0$ and $b_0$
we find the equations
\begin{eqnarray} \label{sb6}
b_0^2 + T\sum_{\epsilon_n} \Tr \left[ \hat {\cal G}(\I \epsilon_n)
 \hat \sigma_z  \right] &=& 0, \\ 
\label{sb7}
b_0 \lambda_0 + b_0T \sum_{\epsilon_n} \Tr \left[ \hat{\cal G}(\I \epsilon_n)
 \hat \Gamma(\I \epsilon_n) \right] &= &0  
,\end{eqnarray}
which have to be solved self-consistently.
$\hat{\cal G}(\I \epsilon_n)$ is the pseudo fermion Green's function
given by
$\hat{\cal G}(\I \epsilon_n) =
[\I \epsilon_n \hat\sigma_0 - \tilde \epsilon_{\rm d} \hat\sigma_z -b_0^2
  \hat \Gamma(\I \epsilon_n)]^{-1}$.
Both in the limit of small and large superconducting gap, we are able to 
solve the mean field equations analytically as demonstrated here below.
The first equation, eq.(\ref{sb6}), is the constraint. 
Since the pseudo fermion level is
at maximum singly occupied, the renormalized level is above the Fermi energy.
In the Kondo limit, where the occupancy is nearly one, we find that 
$0 < \tilde \epsilon_{\rm d} < b_0^2(\Gamma_{\rm N} + \Gamma_{\rm S}) $, 
{\it i.e.} $\lambda_0 \approx | \epsilon_{\rm d} |$ and
$\tilde \epsilon_{\rm d} \approx 0$.
The renormalization of the tunneling amplitude is determined 
from eq.(\ref{sb7}). A trivial
solution $b_0=0$ always exists. The solutions which minimize 
the free energy, however, are those 
with $b_0 \ne 0$. 
By introducing a flat density of states in the leads and the tunneling rates
$\Gamma_\eta= 2\pi N_{0}|V_\eta|^2$, 
the elements of the matrix
$\hat \Gamma (\I \epsilon_n )$ are
$\Gamma_{11}=\Gamma_{22}=-\I\gamma_1$ and
$\Gamma_{12}=\Gamma_{21}^*=\gamma_2$, where
\begin{equation} 
\label{sb8}
\gamma_1  = {\sign}(\epsilon_n) {\Gamma_{\rm N} \over 2} 
+{\Gamma_{\rm S}\over 2} {\epsilon_n \over \sqrt{\epsilon_n^2 + |\Delta|^2} },~ 
\gamma_2  ={ \Gamma_{\rm S} \over 2}
 { \Delta \over \sqrt{\epsilon_n^2 + | \Delta|^2} }
.\end{equation}
Restricting ourselves to zero temperature, we replace the Matsubara 
sum in eq.(\ref{sb7}) by an integral and obtain 
\begin{equation}
\label{sb12}
|\epsilon_{\rm d} | = 4 \int_0^W {\D \epsilon \over 2 \pi } 
{ \gamma_1 ( \epsilon + b_0^2 \gamma_1 ) + b_0^2 | \gamma_2|^2  \over
( \epsilon+ b_0^2 \gamma_1 )^2 + b_0^4 | \gamma_2 |^2}
,\end{equation}
where $W$ is a cut-off of order  the band-width.
We simplify the integral by approximating $\gamma_1 $ and $\gamma_2$ as
\begin{equation}
\label{sb13}
\gamma_1 =  \left\{ \begin{array}{ll}
\Gamma_{\rm N}/2  & {\rm for }\,\,\, \epsilon < \Delta  \\
(\Gamma_{\rm N}+ \Gamma_{\rm S})/2 & {\rm for }\,\,\, \epsilon > \Delta
\end{array} \right. 
,~ \gamma_2 = \left\{
\begin{array}{ll}
\Gamma_{\rm S}/2 & {\rm for} \,\,\, \epsilon < \Delta \\
0        & {\rm for} \,\,\, \epsilon > \Delta 
\end{array} \right.
.\end{equation}
The result is
\begin{eqnarray}
\label{ab15}
|\epsilon_{\rm d}|  &= & {\Gamma_{\rm N} \over 2 \pi }
             \ln{ (2 \Delta + b_0^2 \Gamma_{\rm N})^2 + b_0^4\Gamma_{\rm S}^2 
  \over
                   b_0^4( \Gamma_{\rm N}^2+\Gamma_{\rm S}^2) } \nonumber\\ 
&& +{\Gamma_{\rm N}+\Gamma_{\rm S} \over 2\pi }
\ln{ 4 W^2 \over (2 \Delta +b_0^2 \Gamma_{\rm S} +b_0^2 \Gamma_{\rm N} )^2  }
,\end{eqnarray}
where we neglect a term proportional to $\Gamma_{\rm S}$, but without 
any logarithmic factor.
If $\Delta $ is much smaller than
the Kondo 
temperature which is given by 
$T_{\rm K} = b_0^2  \Gamma_{\rm N}+ b_0^2 \Gamma_{\rm S}$,
$\Delta$ is negligible. One can then easily solve eq.(\ref{ab15}) 
for $b_0^2$ and obtain the
result for two normal leads with total tunneling rate 
$\Gamma_{\rm N}+\Gamma_{\rm S}$: 
\begin{equation}
\label{sb16}
b_0^2(\Gamma_{\rm N}+\Gamma_{\rm S})
 = 2 W \exp\left( -{\pi |\epsilon_{\rm d}|\over \Gamma_{\rm N} + \Gamma_{\rm S} }
\right) 
.\end{equation}
In the opposite limit, where $\Delta$ is much larger than $T_{\rm K}$, we find
\begin{equation}
\label{sb18}
b_0^2 \sqrt{ \Gamma_{\rm N}^2 + \Gamma_{\rm S}^2 } = 2 W \exp \left(  -{\pi }
{ |\epsilon_{\rm d}| - ( \Gamma_{\rm S} /  \pi ) \ln(W / \Delta) \over \Gamma_{\rm N} } \right)
.\end{equation} 
The results agree qualitatively with what we expect from scaling arguments
for the Anderson model.
In the perturbative regime, a logarithmic correction to $\epsilon_{\rm d}$ has been 
found (Haldane 1978). 
This applies to
the case of a large gap, since scaling due to the superconducting electrons stops
at energies of the order $\Delta$, giving rise 
to a logarithmic renormalization of 
$\epsilon_{\rm d}$, as seen in eq.(\ref{sb18}).
In the case of a small gap, the superconducting lead contributes to scaling down 
to low energies, where one enters the strong coupling regime.
Presumably, the fixed point is still reached for energies of the order of 
$T_{\rm K}$, much greater than $\Delta$, so that the Kondo temperature 
does not depend on $\Delta$, as indeed found in eq.(\ref{sb16}).
Notice that in the presence of normal electrons, 
we always find a non-trivial solution of the
mean field equations. This is
to be contrasted with the case of superconducting
electrons only, $\Gamma_{\rm N}=0$, where for large gap
only the solution $b_0=0$ exists, and there is no
Kondo effect (Borkowski and Hirschfeld 1994).

In a non-equilibrium situation, when a voltage is applied between the two leads,
the mean field parameters cannot be obtained by minimizing the free energy.
However the mean field equations (\ref{sb6},\ref{sb7}) 
can also be derived using a self-consistent 
diagrammatic method (Millis and Lee 1987). 
Then it is straightforward to generalize to non-equilibrium.
The equations read
\begin{eqnarray} \label{sb21}
b_0^2 -\I \int {\D \epsilon \over 2\pi } \Tr \left[ \hat{\cal G }^<(\epsilon )
 \hat \sigma_z \right] &=& 0 \\
\label{sb22}
\lambda_0 b_0 -\I b_0 \int{\D\epsilon \over 2 \pi} \Tr \left[ 
\hat {\cal G}^R(\epsilon) \hat \Gamma^<(\epsilon) 
+ \hat{\cal G}^<(\epsilon) \hat \Gamma^A(\epsilon)
\right] &=&0
,\end{eqnarray}
where the lesser Green's function
$ \hat{\cal G}^<(t, t' ) = \I \langle \phi^\dagger (t')
 \phi(t)    \rangle$
has been introduced, with 
$\phi=( f_\uparrow, f^\dagger_\downarrow )$.
The lesser and advanced matrix $\hat \Gamma$ 
is defined in analogy to its equilibrium 
version in eq.(\ref{sb3}).
To obtain $\hat {\cal G}^<$, we use the general relation
$\hat {\cal G}^<= \hat{\cal G}^R \hat \Sigma^< \hat{\cal G}^A$,
where at mean-field level $\hat \Sigma^<= b_0^2 \hat \Gamma^< $ and
\begin{equation}\label{sb23}
{\hat {\Gamma}}^< (\epsilon)=-\sum_{\eta, {\bf k} } 
|V_\eta|^2{\hat\sigma}_z
\left[ {\hat g }_{\eta,{\bf k}}^R(\epsilon) {\hat {f}}_{\eta} (\epsilon) -
 {\hat {f}}_{\eta} (\epsilon){\hat {g}}_{\eta,{\bf k}}^A(\epsilon)\right]
{\hat\sigma}_z.
\label{nonintself}\end{equation}
Note that the superconducting lead does not contribute
to $\Sigma^< (\epsilon )$ for $| \epsilon | < \Delta$. 

We have solved the mean-field equations in the presence of an external 
voltage numerically. As long as $|e V| \ll T_{\rm K}$ the solution 
is almost independent of the voltage. For large voltage, 
$|eV| \gg T_{\rm K}$, we have found that the Kondo peak is pinned to the
chemical potential in the normal lead, {\it i.e.} 
$\tilde\epsilon_{\rm d} \to \tilde\epsilon_{\rm d} -eV$, 
and the peak width is decreased.

The Andreev current can now be determined using the current 
formula eq.(\ref{current4}),
with the dot Green's function 
$\hat{  G }= b_0^2 \hat {\cal G}$. The spectral function, which is
defined as in eq.(\ref{current55}), is determined as
\begin{equation}
\label{conductance}
G_{{\rm NS}}(\epsilon )= {4 e^2 \over h}
{4 {(\tilde\Gamma_{\rm N} \tilde\Gamma_{\rm S} )^2 }\over
(4\tilde \epsilon^2 -4{ \tilde\epsilon}_{\rm d}^2-
\tilde \Gamma_{\rm N}^2-\tilde\Gamma_{\rm S}^2)^2
+ 16  \tilde\Gamma_{\rm N}^2 \tilde \epsilon^2 } 
\end{equation}
Here the renormalized tunneling rates 
$\tilde\Gamma_{{\rm S},{\rm N}}=b_0^2\Gamma_{{\rm S},{\rm N}}$,
and
${ \tilde {\epsilon}} = \epsilon (1+b_0^2 \Gamma_{\rm S}/2\Delta)$ are introduced.
One recovers the current formula for a non-interacting 
quantum dot (Beenakker 1992), 
with renormalized parameters which are voltage dependent.
On resonance,
when $\tilde\epsilon_{\rm d} \approx 0$ and $\epsilon =0$,
the small renormalization factor $b_0$ drops out.
The differential conductance becomes maximal when
$\tilde{\Gamma}_{\rm N}=\tilde{ \Gamma}_{\rm S}$ with 
$G_{{\rm NS},{\rm max}}=4e^2/h$,
twice the maximum for a N-QD-N system.
For large voltage $G_{\rm NS}$ drops quickly, 
since the resonance moves away from zero energy, 
$|\tilde\epsilon_{\rm d}  | \approx  |eV|$. 
$G_{\rm NS}$ as a function of energy is proportional
to $(eV)^{-2}$ near $\tilde \epsilon \approx \pm \tilde \epsilon_{\rm d} $ 
and proportional
to $(eV)^{-4}$ near $\epsilon = 0$.
As a consequence the current decreases with increasing voltage, 
leading to a negative differential conductance. 
This is also demonstrated in Fig.~\ref{fig4}, where 
the current as a function of voltage obtained 
using $G_{\rm NS}$ of eq.(\ref{conductance}) with the numerically determined,
voltage dependent mean field parameters.
The results were obtained with
$\epsilon_{\rm d }= W/3$, $\Gamma_{\rm N}=\Gamma_{\rm S}=0.14W$, and $\Delta = 0.2W$. 
For low voltages the
current is, to good approximation, given by 
$I= 4 e^2 V/ h$, whereas the current drops when the voltage exceeds the Kondo temperature.

Finally, we want to comment on the reliability of our results. 
The success of slave-boson mean field theory stems from the fact
that it captures the  Fermi-liquid regime at low temperature. 
If the N-QD-S system  scales to a Fermi liquid 
at low temperature, $G_{\rm NS}$ as given in eq.(\ref{conductance}) is
exact in the low
temperature, low voltage limit.
Since it is known that slave boson mean field theory has problems in describing
dynamical properties, the results far away from equilibrium need 
to be treated with caution.

Within the Fermi-liquid point of view, the present mean field approach allows
us to 
estimate the parameters entering eq.(\ref{conductance}).
In particular, we found that $\Gamma_{\rm N}$ and $\Gamma_{\rm S}$ 
renormalize equally, but this may no longer be the case when
considering higher order corrections.
For illustration, we estimate the effect of residual quasiparticle interaction
in the limit $\Delta \ll T_{\rm K}$. By assuming an effective quasiparticle
interaction of the form
$H_{\rm int} = \tilde U n_\uparrow n_\downarrow$, we find to first order
in $\tilde U$ no corrections to $\tilde{\Gamma}_{\rm N}$, while, 
as one could have expected, repulsive quasiparticle interaction 
suppresses the renormalized coupling to the superconductor, 
$\tilde{\Gamma}_{\rm S} =b_0^2{\Gamma}_{\rm S}[1-
4(\tilde U /\pi T_{\rm K})(\Delta / T_{\rm K})\ln{T_{\rm K}/2 \Delta}]$.

\section{Conclusions}
We have studied Andreev tunneling in a normal metal-quantum dot-superconductor device, and 
obtained a general formula for the current through the device.
We calculated the current voltage characteristics within the 
equation-of-motion approach, where we found a suppression 
 of the conductance at low 
temperature. However it is known that this approach does not provide 
quantitative results in the Kondo regime. Therefore we extended the 
analysis to the extreme low temperature regime  using slave boson mean 
field theory. In the regime when the superconducting gap is smaller than
the Kondo temperature, we found 
an enhanced Andreev current at low bias voltage due to the Kondo effect. 
The zero bias conductance is maximum with the universal value 
$G_{\rm NS}=4e^2/h$ 
when the renormalized tunneling rates $\tilde \Gamma_{\rm N}$ and $\tilde 
\Gamma_{\rm S}$ are equal. 
We identified the ratio 
$\Delta/T_{\rm K}$ as an important parameter.
In the case $\Delta \ll T_{\rm K}$,  
the Kondo resonance forms as for two normal leads.
The condition $\tilde \Gamma_{\rm N} = \tilde \Gamma_{\rm S}$ coincides with 
equal bare tunneling rates ${\Gamma}_{\rm N}={\Gamma}_{\rm S}$. In the case of
large gap, quasiparticle interaction suppresses $\tilde{\Gamma}_{\rm S}$,
nevertheless the conductance maximum condition may be achieved
in an asymmetric QD where ${\Gamma}_{\rm S} > {\Gamma}_{\rm N}$.

\acknowledgements
We thank R. Fazio for many fruitful discussions.
We acknowledge the financial support of INFM under the PRA-project
"Quantum Transport in Mesoscopic Devices" and the EU TMR programme. 
We acknowledge a useful correspondence with A. Clerk
which made us aware of a mistake in our original manuscript and in 
(Fazio and Raimondi 1998).

{\epsfxsize=7cm\epsfysize=5cm\epsfbox{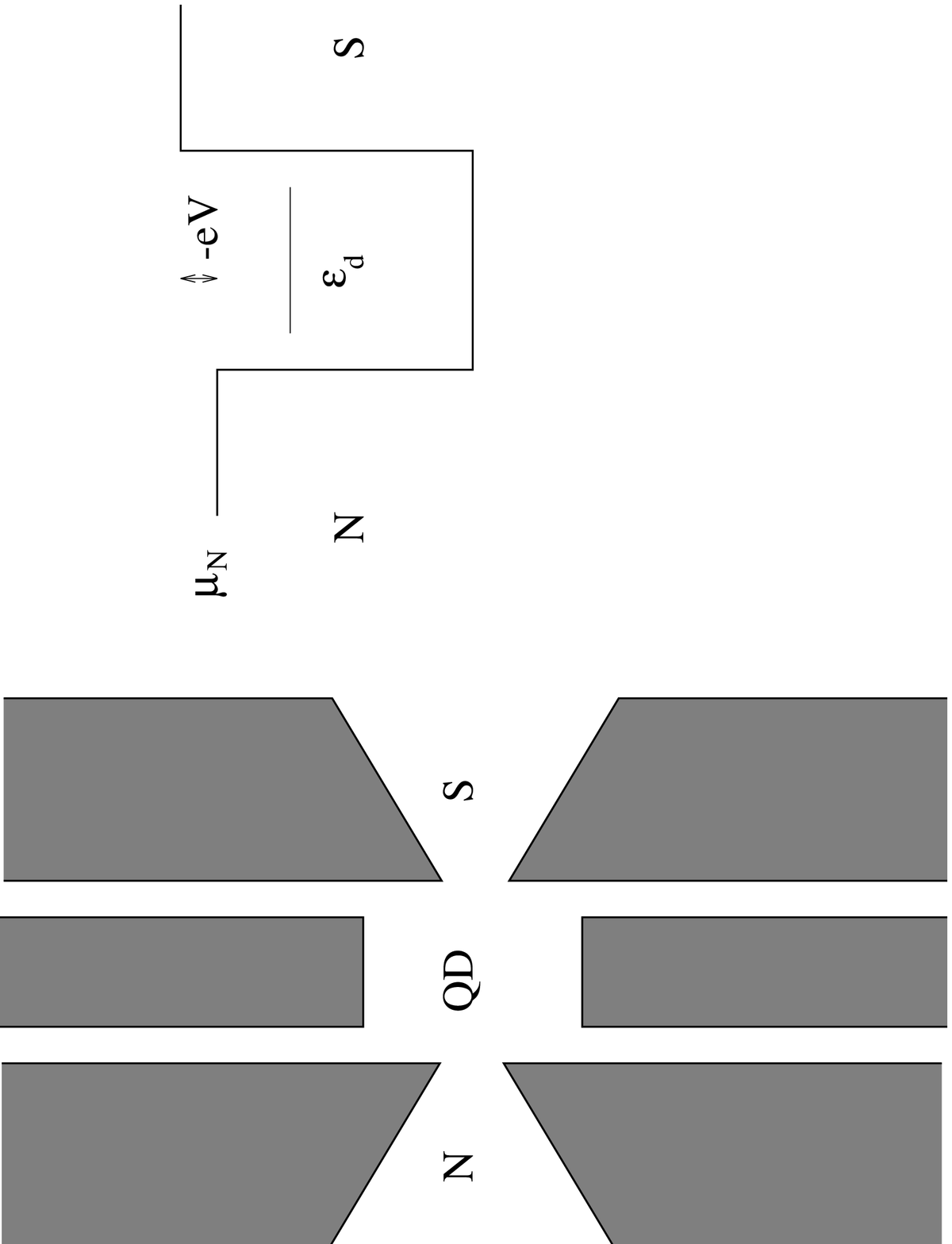}}
\begin{figure}
\narrowtext
\caption{The system under consideration. A quantum dot coupled by tunnel 
barriers to a normal and to a superconducting electrode. The position of the
level in the dots can be tuned by means of the gate voltage $V_g$}
\label{fig_bag_dot}
\end{figure}

{\epsfxsize=7cm\epsfysize=7cm\epsfbox{fig2.pstex}}
\begin{figure}
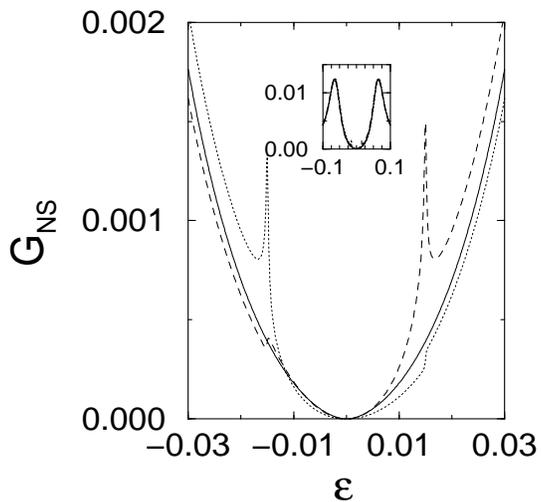

\caption{\narrowtext
The spectral density for the two particle 
tunneling $G_{\rm NS}(\epsilon)$  is plotted for various bias voltages ( 
$V=0$ solid line, $ V=- 0.015$ dotted line, 
$ V=0.015$ dashed line,  
$\tilde{\epsilon}_{\rm d} = -0.07$, 
$\Gamma_{\rm S}=\Gamma_{\rm N} = 0.02$ and $T = 0.0001$, 
$\Delta =0.1$, in units
of the bandwidth $W$). 
In the inset the same curves are shown in an extended scale.}
\label{fig2}
\end{figure}
{\epsfxsize=7cm\epsfysize=7cm\epsfbox{fig3.pstex}}
\begin{figure}
\caption{\narrowtext 
The differential conductance of the N-QD-S device, in units of
$4e^2/h$, is plotted for 
different temperatures ($T=0.0001$ solid line, $ T = 0.001$ dotted line, 
$ T= 0.01$ dot-dashed line, $\tilde{\epsilon_{\rm d}} = -0.04$, 
$\Gamma_{\rm S}=\Gamma_{\rm N} = 0.02$, $\Delta =0.1$, in units
of the bandwidth $W$).}
\label{fig3}
\end{figure}
{\epsfxsize=7cm\epsfysize=7cm\epsfbox{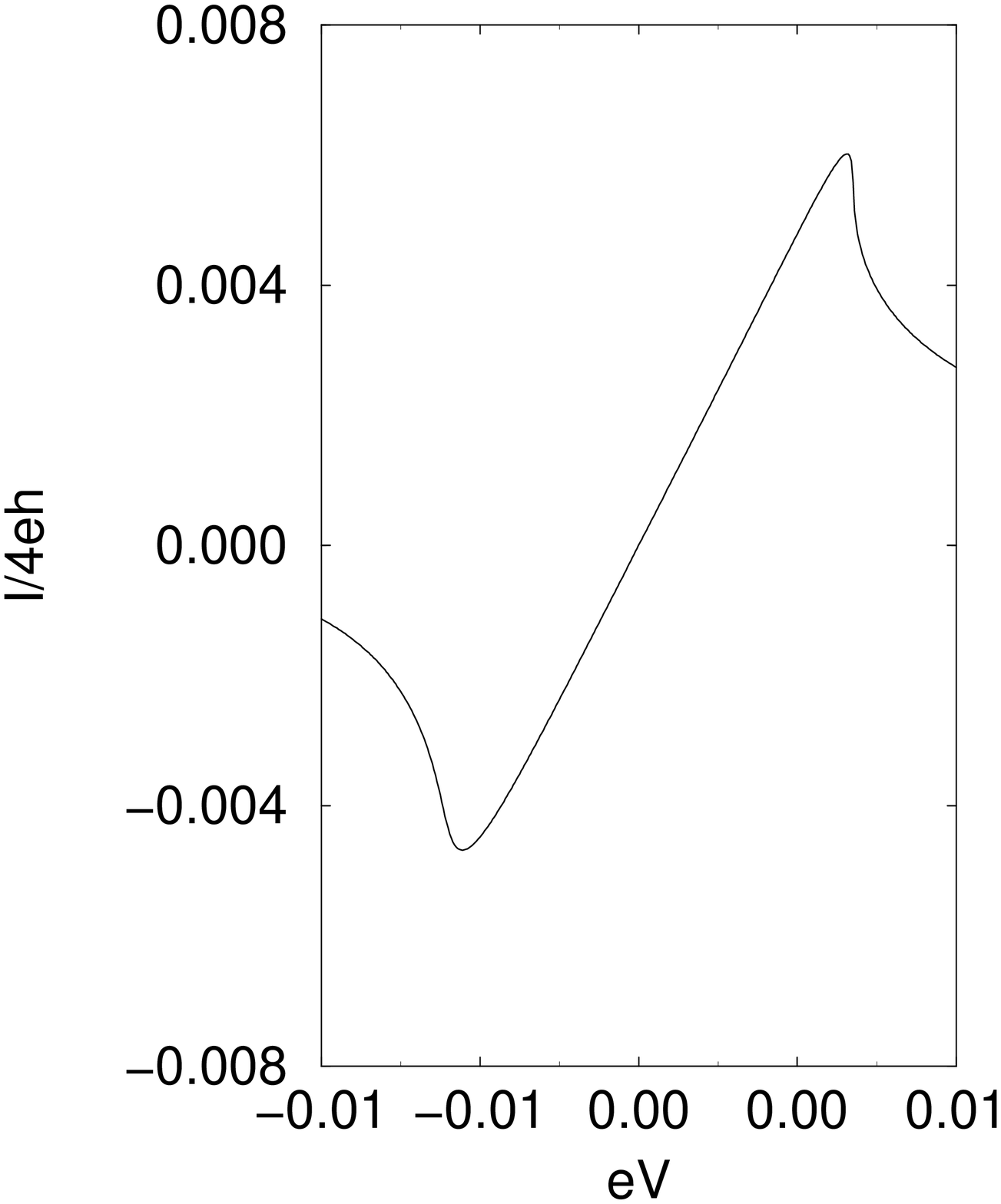}}
\begin{figure}
\caption{
\narrowtext
Current voltage characteristics of a quantum dot at zero temperature
as obtained within slave boson mean field theory. 
$\epsilon_{\rm d} =1/3$, $\Gamma_{\rm N}= \Gamma_{\rm S}=0.14$, and $\Delta =0.2$
in units of the bandwidth.
The Kondo temperature in equilibrium is 
$T_{\rm K} \approx  b_0^2 \Gamma_{\rm N} =0.014$.}
\label{fig4}
\end{figure}
\end{multicols}
\end{document}